\documentclass{osa-article}

\journal{oe}

\usepackage{siunitx}
\usepackage{graphicx}

\usepackage{xcolor}

\begin{document}

\title{A simple, pixel-wise response correction for ring artifact removal in both absorption and phase contrast X-ray computed tomography}

\author{Linda C. P. Croton,\authormark{1,*} Gary Ruben,\authormark{1}, Kaye S. Morgan\authormark{1,2}, David M. Paganin\authormark{1}, and Marcus J. Kitchen\authormark{1}}

\address{\authormark{1}School of Physics \& Astronomy, Monash University, Victoria 3800, Australia\\
\authormark{2}Chair of Biomedical Physics, Department of Physics, Munich School of Bioengineering, and Institute of Advanced Study, Technical University of Munich, 85748 Garching, Germany}

\email{\authormark{*}linda.croton@monash.edu} 



\begin{abstract*}
We present a pixel-specific, measurement-driven correction that effectively minimizes errors in detector response that give rise to the ring artifacts commonly seen in X-ray computed tomography (CT) scans. This correction is easy to implement, suppresses CT artifacts significantly, and is effective enough for use with both absorption and phase contrast imaging. It can be used as a standalone correction or in conjunction with existing ring artifact removal algorithms to further improve image quality. We validate this method using two X-ray CT data sets, showing post-correction signal-to-noise increases of up to 55\%, and we define an image quality metric to use specifically for the assessment of ring artifact suppression.\\
\end{abstract*}

\hrule
\bibliography{refabbrev}
\hrule

\section{Introduction}

Ring artifacts occur in X-ray computed tomography (CT) due to an amalgamation of factors that cause small errors in detector pixel values to persist throughout CT acquisition, resulting in semi-circular and ring-shaped artifacts on back projection for 180$^\circ$ and 360$^\circ$ CTs, respectively. Many of these errors are caused by phenomena that affect the detector gain, such as phosphor thickness variations or imperfections in the optical coupling system, while others include fluctuations in the X-ray beam and, for synchrotron sources, drift and/or vibrations in the monochromator crystal. Were the X-ray beam perfectly stable for the duration of the CT and the detector response uniformly linear for every pixel, standard dark-current and flat-field correction techniques would be sufficient to correct for these inhomogeneities and remove these artifacts; however, since detector gain and dark current can vary considerably in space and time, standard flat and dark correction is generally inadequate to compensate for these intensity variations across the entire dynamic range. Therefore, other techniques are required for proper treatment of these artifacts. 

A number of different methods have been proposed for ring artifact correction, most of which are based on post hoc image processing. Many of them isolate and remove rings in the reconstructed image \cite{Jha2014, Paleo2015, Ji2017, Liang2017} or in the sinogram \cite{Raven1998, Sijbers2004, Boin2006, Muench2009, Yousuf2010, Miqueles2014, Titarenko2016, Yan2016, Massimi2018, Vo2018}. Some methods characterize the flat-field images \cite{VanNieuwenhove2015, Jailin2017}, while others shift the sample or detector during image acquisition to smear out systematic intensity fluctuations across the reconstruction volume \cite{Davis1997,Hubert2018,Pelt2018}. Of particular interest are those that seek to correct the bulk of the problem where it occurs -- in the pixel-to-pixel response variations that have historically been treated as spatially invariant \cite{Altunbas2014, Vagberg2017, Vo2018}. While all of these can be effective at addressing the ring artifacts, we take the latter approach, since it is directed at the root cause and is therefore likely to provide more accurate results without leading to other artifacts, such as blurring or the introduction of new rings. V{\aa}gberg~\textit{et~al.}~(2017)~\cite{Vagberg2017} recently took a similar approach, proposing a measurement-informed ring artifact correction algorithm, modeling the detector response using images collected over a range of intensities.  Their work used aluminum filters to attenuate the beam and took the assumption that the spatial variations in response are primarily caused by changes in thickness across the scintillator. The algorithm presented here is effectively a generalization of this work, making no assumptions regarding the cause of the detector's spatial variations, using a method that is sample-independent. We present a simple, pixel-wise detector calibration using hundreds of data points for each position on the detector, rather than the standard two-point flat-field calibration. This method requires only a single image sequence for each experimental setup and does not require any information about the physical cause of the detector's response variations.

This approach can be used for conventional X-ray CT but was primarily motivated for the case of phase-contrast X-ray CT \cite{Langer2008}.  Phase contrast is a relatively recent development, which converts the phase shifts imparted to the X-ray wavefield by the sample into intensity variations that can reveal soft tissue features. The simplest way to achieve this additional soft-tissue contrast is through the introduction of a distance (e.g.~>~1~m) between the sample and detector, illuminating the sample with sufficiently coherent X-ray radiation \cite{Snigirev1995,Cloetens1996}.  Since phase-contrast X-ray imaging can capture soft tissue structures, we have recently applied the method to image the brain \textit{in situ}, finding that the phase contrast associated with grey and white matter is orders of magnitude less than the attenuation contrast associated with the skull \cite{Croton2018}. The correction presented here is therefore of particular interest in phase contrast imaging, given that the image contrast associated with phase effects may be comparable to the contrast of ring artefacts.

\section{Pixel-wise mapping of the detector response}
\label{sec:maps}

The full correction can be broken into two main parts -- first, a spatial mapping of the detector response, followed by an application of that mapping to experimental data. In the following sections, we break these parts further into eight basic steps, labeled (1) - (8) below. To determine the spatial variations in response across the detector, we only need to know how the intensity measured by the optical system at each pixel varies from that which is incident upon it. These intensity deviations are not expected to have a large dependence on energy, since the incident X-ray photons are converted to optical photons within the system; however, there may be some dependence if the point spread function changes significantly with energy. Here, we minimize energy dependence by measuring the response at similar energies to those used for our experiments, using a monochromatic source. We take a series of measurements acquired across a range of intensities covering the bulk of the dynamic range of the detector. It is possible to use filters to attenuate the beam \cite{Vagberg2017}, however spatial variations in density and thickness of the filters can add artifacts back into the images.

\begin{figure}[h!]
  \centering
      \includegraphics[width=\textwidth]{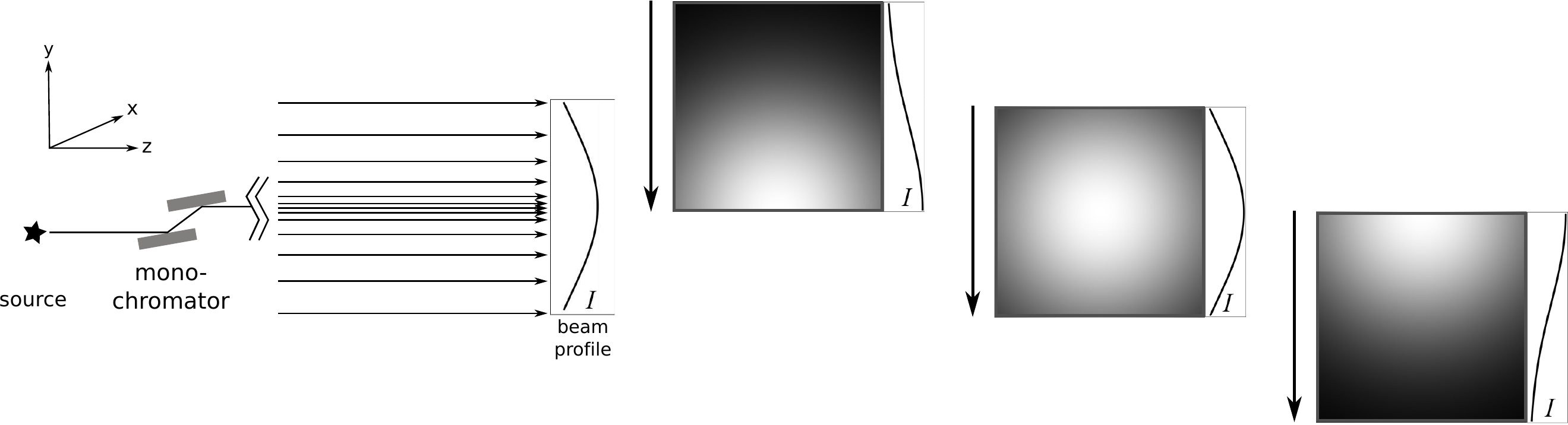}
    \caption{The experimental set-up, shown for an incident beam with a parabolic intensity profile. When sweeping vertically, the detector is positioned high enough above the beam to achieve the desired minimum number of incident counts.  A sequence of images is then acquired while the detector is translated downward through the beam to the equivalent position below the beam. Three images are shown near the middle of the sequence, along with intensity profiles taken vertically through the center of the images. The extremes ends of the sequence (not shown) contain only traces of the edge of the incident beam. Note that the direction of transverse translation is not important; it is equally valid to acquire an image sequence while translating the detector horizontally through the beam.}
  \label{fig:setup}
\end{figure}

Step 1 -- To avoid introducing new artifacts, we start by simply sweeping the detector through the X-ray beam while acquiring a sequence of images (see Fig. \ref{fig:setup}). Since the beam intensity in the absence of any sample is typically peaked at the center, rolling off outward away from the beam, a wide range of intensity measurements can be obtained. For a beam size that is larger than the detector, a single sweep is usually sufficient; when the detector is larger, more than one sweep with the beam offset from the center of the detector may be necessary to ensure coverage across the full range of intensities for every pixel. By eliminating the need for absorbing materials, we can measure the exact response for a given input intensity, thus avoiding any possible artifacts that may be introduced by structural imperfections in the absorber. If, however, the intensity profile of the beam is insufficient to sample the full range of intensities required, then filters can be added to extend the intensity range acquired; the structural imperfections would not introduce additional artifacts when used in this way, since they would be shifted across the extent of the detector in the sweep direction during acquisition.

Image sequences were acquired on beamline BL20B2 at the SPring-8 synchrotron in Hy$\overline{\mbox{o}}$go, Japan using a 2048 $\times$ 2048 ORCA Flash 4.0 digital sCMOS camera (C11440-22C by Hamamatsu) with a \SI{25}{\um} thick gadolinium oxysulfide (GOS) scintillator coupled with a tandem lens system, giving an effective pixel size of \SI{15.1}{\um}. Three beam sweeps were required for full coverage across the detector. 270 images were acquired in each sequence, with an exposure time of 100~ms each. Sequences were recorded at three positions such that the beam was centered on the horizontal left, center, and right side of the detector, which was swept vertically.

\begin{figure}[h!]
  \centering
      \includegraphics[width=\textwidth]{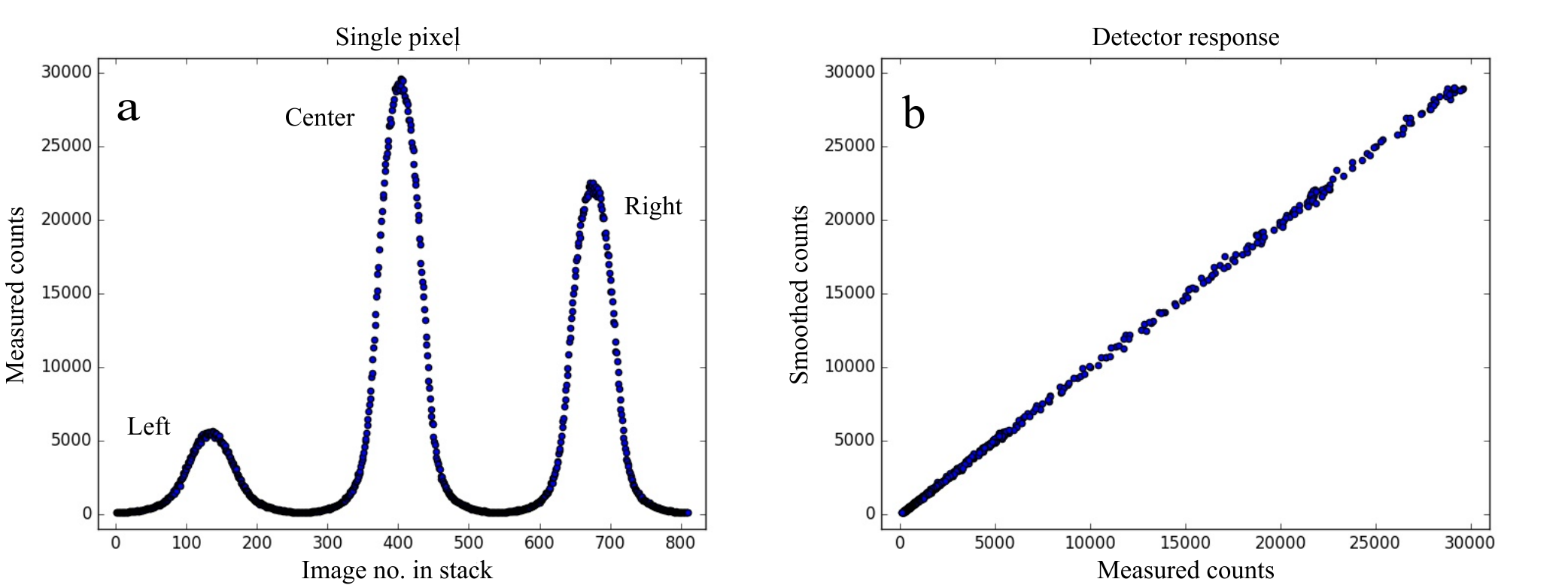}
    \caption{a) The counts measured at a single pixel in a stack of images taken while the detector was swept vertically through the beam three times; the beam was centered first on the horizontal left, then center, and finally right sides that this pixel is near the horizontal center. b) The `true' (i.e. smoothed) counts for the same pixel as in (a) as a function of the measured counts.}
  \label{fig:plots}
\end{figure}

Step 2 -- Once the full beam sweep sequence was acquired, the images were stacked into a volume $I(i,j,k)$, where $i$ and $j$ are the spatial dimensions within the $k$ images in the stack. Figure \ref{fig:plots}(a) shows the intensity through the combined image stack for a single pixel.

Step 3 -- A separate volume was created wherein each image $I(i,j)$ of the volume $I(i,j,k)$ was first dark-corrected using the mean image $D(i,j)$ of the dark-current images acquired at the time of the sequence, and then smoothed with a Gaussian filter, yielding an estimate $I_s(i,j,k)$ of the true intensity profile of the incident X-ray beam. The smoothing kernel radius was chosen to be large enough to eliminate the pixel-to-pixel variations resulting from the non-uniform gain, while being small enough to maintain the underlying shape of the beam intensity profile. We found that a kernel with a standard deviation of 50 pixels provided a suitable blurring function for our experimental conditions.

Step 4 -- The `true' beam intensity was plotted against the measured intensity to determine the order of polynomial, for smoothed counts as a function of measured counts, required to model the gain. This is shown in Fig. \ref{fig:plots}(b). Since the two intensities come from the same measurement, this resultant curve is necessarily linear, with a slope close to unity.

Step 5 -- The `true' intensity $I_s(i,j,k)$ was fit for each pixel $(i,j)$ as a linear function of the measured intensity,
\begin{equation}
I_s(i,j,k) = \boldsymbol\alpha [I(i,j,k) - D(i,j)] + \boldsymbol\beta.
\label{eq:linearfit}
\end{equation}
The coefficient arrays $\boldsymbol\alpha = \alpha(i,j)$ and $\boldsymbol\beta = \beta(i,j)$ from this calibration (specified in bold for clarity) can then be used to determine the `true' intensity of the beam $I_s(i,j)$ incident on the detector, given the measured intensity $I(i,j)$ for a given projection image, regardless of the sample. It should be noted that this will correct for spatial variations in the detector gain and offset, which are those primarily responsible for ring artifacts; however, since $I_s$ is estimated by smoothing the data itself, this method cannot account for any large-scale non-linearities (e.g. due to higher-order harmonics or polychromaticity); these must be accounted for separately. This, however, should not have a substantial impact on ring artifacts, since they are most prominent when caused by the small-scale, pixel-to-pixel variations.

\begin{figure}[h!]
  \centering
      \includegraphics[width=\textwidth]{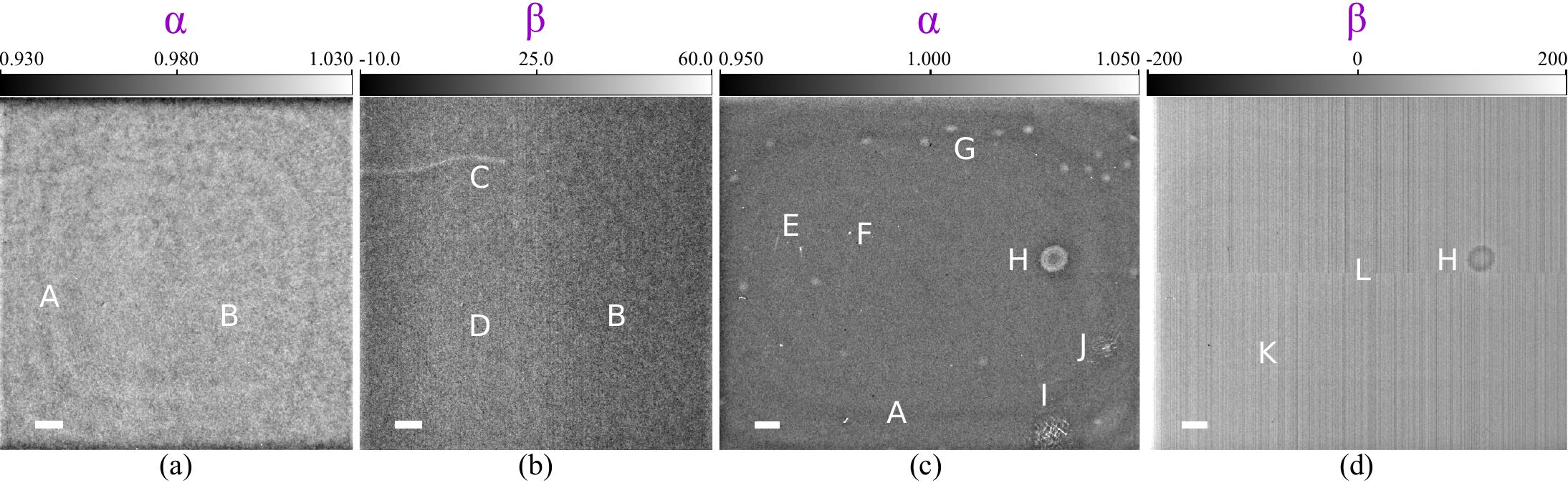}
    \caption{The detector gain ($\boldsymbol\alpha$) and offset ($\boldsymbol\beta$) maps are shown in (a) and (b), respectively, for the Hamamatsu detector used at SPring-8. (c) and (d) show those for the pco.edge detector used at IMBL. Spatial scale bars are 1~mm in length. Numerous features can be seen to affect the gain. These are labeled with white letters and are described at the end of section \ref{sec:maps}.}
  \label{fig:coeffs}
\end{figure}

The coefficient maps, $\boldsymbol\alpha$ and $\boldsymbol\beta$, are shown in Figs. \ref{fig:coeffs}(a) and \ref{fig:coeffs}(b), respectively. Note that $\boldsymbol\alpha$ corresponds to the detector gain, while $\boldsymbol\beta$ maps the intercepts of the fits and hence an offset to the dark current. A distinct line can be seen clearly across the upper left of Fig. \ref{fig:coeffs}(b), and to a lesser degree in Fig. \ref{fig:coeffs}(a), possibly -- though not necessarily -- due to a scratch on the scintillator, as per the assumption of scintillator thickness variations of V{\aa}gberg \textit{et al.} (2017) \cite{Vagberg2017}. Our correction does not require any assumption about the cause of these gain variations, so while effects resulting from thickness variations in the scintillator may be present, any other phenomena affecting the gain would be represented as well.

To test our method under different experimental conditions, a calibration was also performed for a different detector at the Imaging and Medical Beamline (IMBL) of the Australian Synchrotron. Image sequences were acquired using a 2560~$\times$~2160 pco.edge~5.5 sCMOS camera with a tandem lens configuration and a \SI{25}{\um} thick GOS scintillator, giving an effective pixel size of \SI{16.2}{\um}, at a beam energy of 25~keV. Due to the relatively small beam height and narrow vertical slit aperture, five separate image sequences of 300 exposures each were acquired while sweeping the detector across the beam horizontally, rather than vertically, with each sweep centered at a different vertical position on the detector. The images in each sequence were segmented into five strips of equal height, covering the full width of the detector. The strips from each sweep corresponding to the region of peak intensity were combined to form a single final image sequence. This was done to provide a single beam sweep sequence with full coverage across the intensity range at every pixel, while removing the vertical slit edges present in the individual sequences. From this combined sequence, smoothed and unsmoothed volumes were created and processed following the same method as above, yielding the gain and dark-current offset correction maps shown in Figs. \ref{fig:coeffs}(c) and \ref{fig:coeffs}(d), respectively.

Note that there are a number of non-uniformities present in all of the maps in Fig. \ref{fig:coeffs}, of both known and unknown origin. Some features of note include: (A) Dark, concentric bands just inside the perimeters of the detectors. (B) Textured gain variations. (C) A distinct, nearly horizontal line (the aforementioned `scratch'). (D) A broad, vertical, bright band. (E) Nearly parallel, fine bright lines. (F) Small regions of `hot pixels'. (G) Diffuse smudges of unknown origin. (H) A very distinct bright spot on the gain map of the pco.edge detector, also seen as a dark spot in the dark-current offset. This feature results from a small droplet of moisture that fell onto the scintillator during installation. The droplet evaporated quickly, however its effect can be seen in the raw images in addition to the gain and offset maps. (I) and (J) Large, textured `smudges'. (K) Vertical banding corresponding to the readout columns. (L) A horizontal line separating the two vertical readout directions. These features demonstrate that a wide variety of phenomena, even the relatively minor environmental effect of the moisture droplet (H), can substantially influence the detector response.

\section{Applying the correction}
\label{sec:correction}

Step 6 -- To implement the correction for a CT sequence, Eq.~\ref{eq:linearfit} is applied using the coefficient arrays from the beam sweep and replacing the beam sweep volume $I(i,j,k)$ with the projection images $P(i,j,k)$ and using the dark current acquired during the CT sequence, yielding the corrected projections:
 \begin{equation}
P_c(i,j,k) = \boldsymbol\alpha [P(i,j,k) - D(i,j)] + \boldsymbol\beta.
\label{eq:correction}
\end{equation}

Step 7 -- A new flat-field image is created from the mean of the flat-field images acquired with the CT sequence, smoothing the result using the same smoothing parameters as those used to create the beam sweep volume $I_s$.

\begin{figure}[h!]
  \centering
      \includegraphics[width=\textwidth]{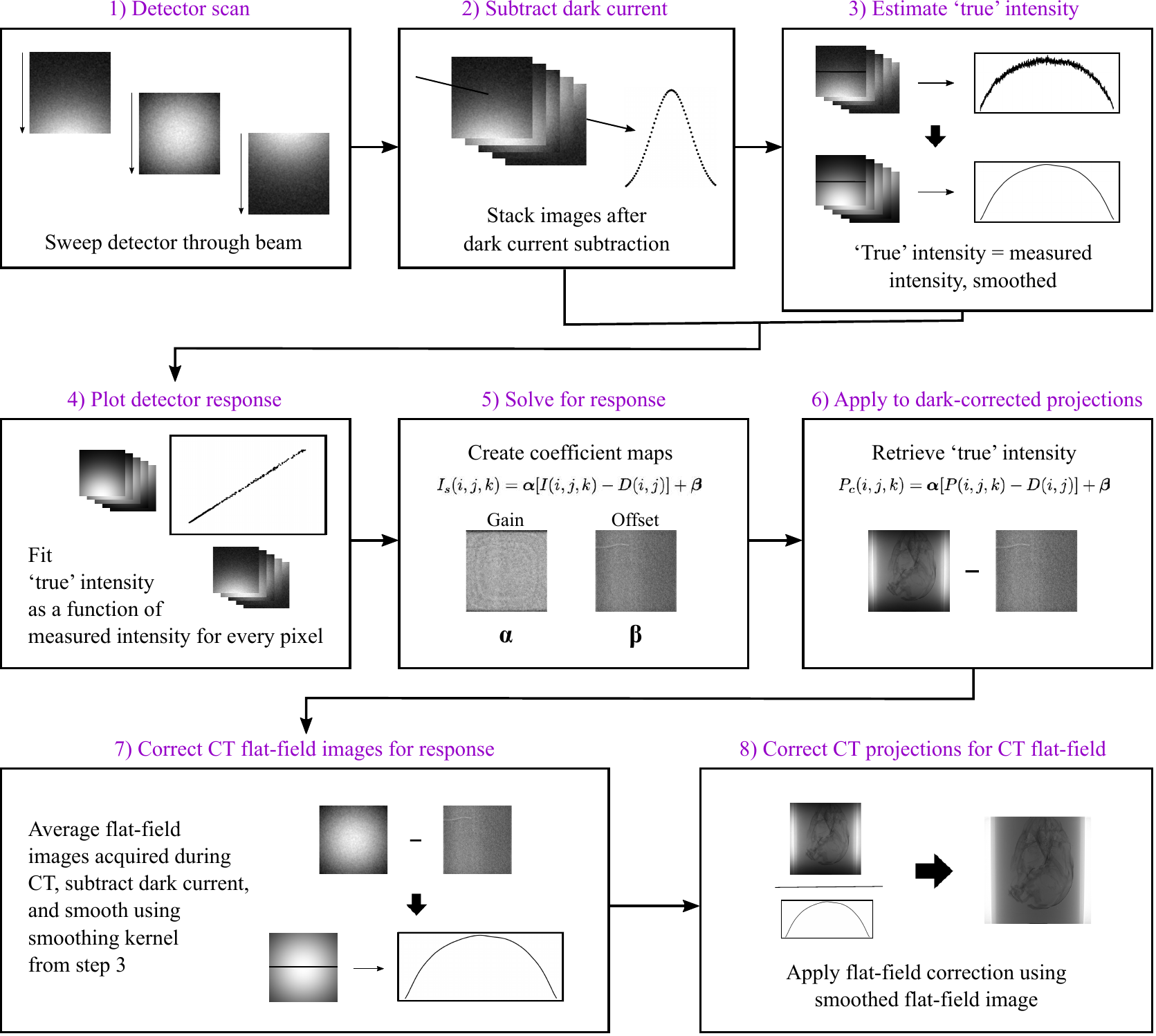}
    \caption{A flow chart of the correction algorithm. The basic method is summarized as follows: (1) The detector is swept through the X-ray beam while an image sequence is acquired. (2) The mean of the dark-current images acquired at the time of the beam sweep is subtracted from each image in the sequence, and they are stacked into a 3D volume. (3) A second volume is created to estimate the `true' intensity of the beam by smoothing each image in the volume. (4) and (5) The measured intensity is fit as a function of the `true' intensity for every pixel, yielding detector gain and dark-current offset maps. (6) For a given CT data set, corrected projections are created using these maps, after subtraction of the mean dark-current images acquired with the CT sequence. (7) The mean dark current is subtracted from the mean of the flat-field images acquired at the time of the CT, and the resultant image is smoothed using the same smoothing kernel as that used in step 3. (8) The corrected projection images are then flat-field corrected using the smoothed flat-field image. See Visualization 1 for a second flow chart showing the additional corrections described in the text.}
  \label{fig:flowchart}
\end{figure}

Step 8 -- Finally, the corrected projection images are flat-field corrected using the smoothed flat-field image. Each of the eight steps is outlined in the flow chart of Fig. \ref{fig:flowchart}.

An additional correction may be required to account for small differences in the incident beam or in the optical system output between the time of the beam sweep and that of the CT data set. This is done by applying the same correction above (Eq.~\ref{eq:correction}) to the mean flat-field image, with residuals calculated as the difference between the corrected and uncorrected images. These residuals are then added to the smoothed flat-field image used for the correction. Additionally, changes in the beam intensity over the course of the CT acquisition due to synchrotron beam injections can be corrected by creating a customized flat-field image for each projection. One simple way to do this is to scale the smoothed flat-field image by the ratio of a reference region in the projections outside the sample to that same region of the mean flat-field image. Visualization 1 contains a flow chart that includes corrections for these residuals and beam injections.

It should be noted that the residual correction is dependent on the amount of noise within the data set. There will likely be unwanted signal within the residuals (e.g. zingers, pixels where the intensity is far from the intensity in surrounding pixels), and when the noise level is low, adding these residuals to the smoothed flat-field image can introduce ring artifacts. In this case, it is important to filter the residuals to ensure that only those that are persistent throughout the CT are included. This can be done by limiting the inclusion of residuals only beyond a certain tolerance (we use those $>3\sigma$ from the mean). When the data set is noisy, this filtering is less important and can even remove residuals that should be included.

\section{Experiments}

During the synchrotron experiments described in the previous section, propagation-based, phase-contrast CTs were acquired of biological samples under the same experimental conditions as those for the beam sweeps.  A CT sequence was acquired at 24~keV at SPring-8 of a scavenged head from a New Zealand White rabbit kitten born at 30 days gestational age (GA; term $\sim$32 days), suspended in agarose. 1800 projections were acquired at a 5~m sample-to-detector propagation distance over 180$^{\circ}$, with an exposure time of 100~ms per projection. A second phase-contrast CT sequence was acquired at the Australian Synchrotron of the lungs of a New Zealand White rabbit kitten, also born at 30 days GA, at 25 keV using a MICROFIL{\textregistered} contrast agent. 3600 projections were acquired at 2~m propagation over 180$^{\circ}$, with an exposure time of 200~ms per projection. To evaluate the effectiveness of our correction, each data set was processed twice, once with traditional dark-current and flat-field corrections and once with the pixel-wise correction detailed in this paper. Phase retrieval was performed using the two-material algorithm derived by Beltran~\textit{et~al.}~(2010)~\cite{Beltran2010} from the single-material algorithm of Paganin~\textit{et~al.}~(2002)~\cite{Paganin2002} and described for CT by Croton~\textit{et~al.}~(2018)~\cite{Croton2018} (for more information, see \cite{Beltran2011,Nesterets2014,Gureyev2014,Kitchen2017,Gureyev2017}). For the head data set, the bone and soft tissue interface was retrieved, while for the lung data set, phase retrieval was performed with respect to the MICROFIL{\textregistered}/tissue interface. In total, four volumes were reconstructed for each data set -- before/after correction without phase retrieval and before/after correction with phase retrieval.
\begin{figure}[h!]
  \centering
      \includegraphics[width=\textwidth]{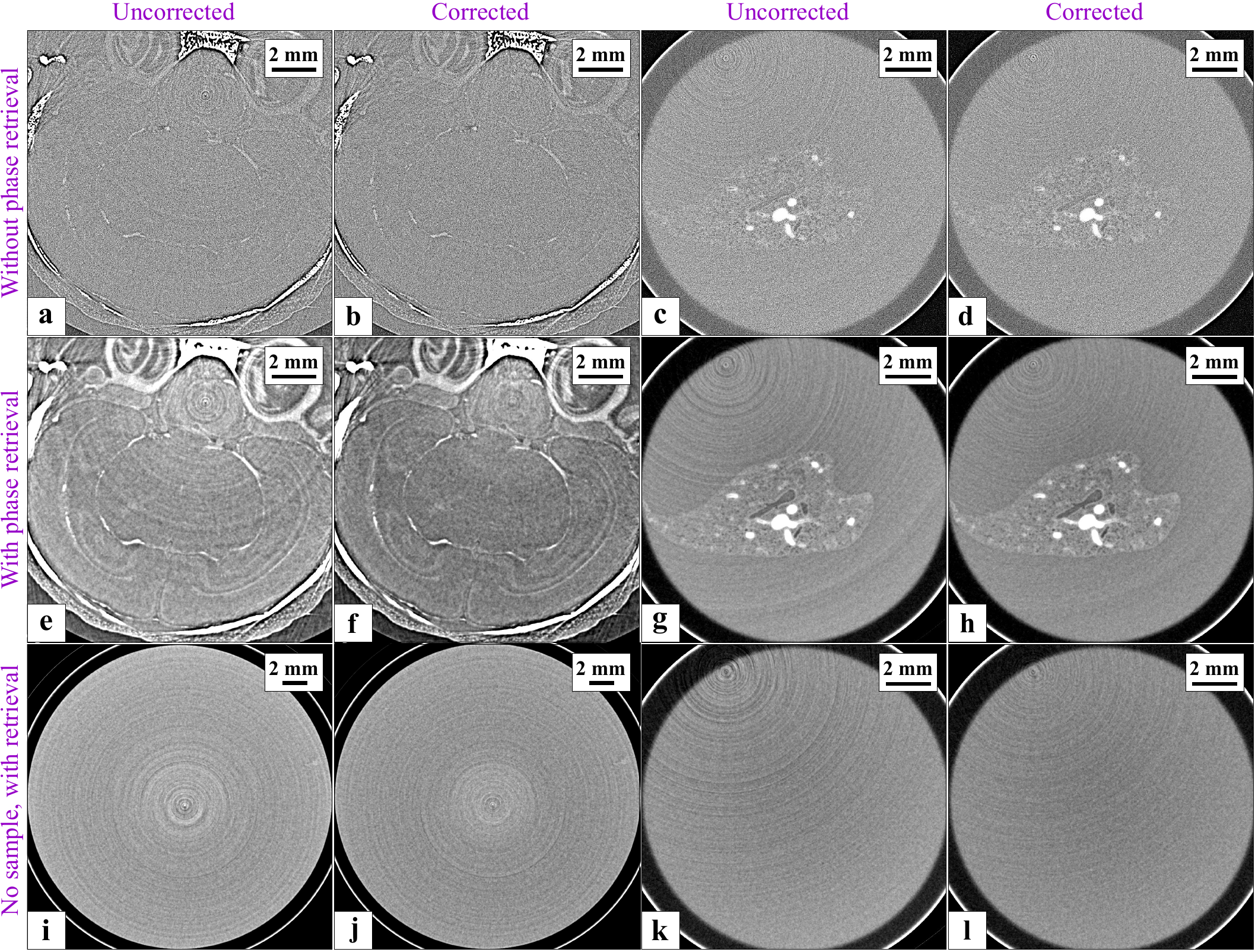}
    \caption{(a), (b), (e), (f) Reconstructed tomograms of a rabbit kitten head in agarose and (c), (d), (g), (h) rabbit kitten lungs in agarose with MICROFIL{\textregistered} contrast agent. Also shown are reconstructed slices of sample-free regions from both the (i), (j)  head and (k), (l) lung data sets, containing only agarose. All data are shown without (first and third columns) and with (second and fourth columns) the correction detailed in this paper. Top row: Phase contrast tomograms of rabbit kitten head and lungs, no phase retrieval. Middle row: Tomograms of the same rabbit kitten head and lungs, after two-material phase retrieval \cite{Beltran2010, Croton2018}. Bottom row: Phase-retrieved tomograms of sample-free regions from rabbit kitten head and lung data sets. First and third columns: Standard dark-current and flat-field correction. Second and fourth columns: Beam sweep gain and offset correction.}
  \label{fig:rings}
\end{figure} 

\section{Results}

Figures \ref{fig:rings}(a)~-~\ref{fig:rings}(l) show reconstructed slices of the rabbit kitten head and lung CTs, both before and after phase retrieval, for standard dark-current and flat-field corrections and for the pixel-wise dark-current offset and gain correction described in section \ref{sec:correction}. Slices were also reconstructed for each data set, post-phase-retrieval, of a sample-free region containing only the sample container filled with agarose. In each of the data sets, the signal-to-noise ratio (SNR), measured as the ratio of the mean to the standard deviation (SNR~=~$\mu / \sigma$) \cite{Smith1997} within the region of interest, increased significantly in the area immediately surrounding the center of rotation (COR) after the correction was applied. We refer the reader to references \cite{Beltran2011,Nesterets2014,Gureyev2014,Kitchen2017,Gureyev2017} for further details regarding this SNR boost. The artifacts are strongest near the COR and fall off with increasing radius, since incorrect pixel values are spread across increasingly larger circumferences. In the innermost region, within a radius of 100 pixels from the COR, SNR increases of 38\% and 39\% were achieved with the correction for the phase-retrieved head and lung images shown in Fig. \ref{fig:rings}, respectively, and increases of up to 55\% were seen across the full sample-free volumes.

\begin{figure}[h!]
  \centering
      \includegraphics[width=1.0\textwidth]{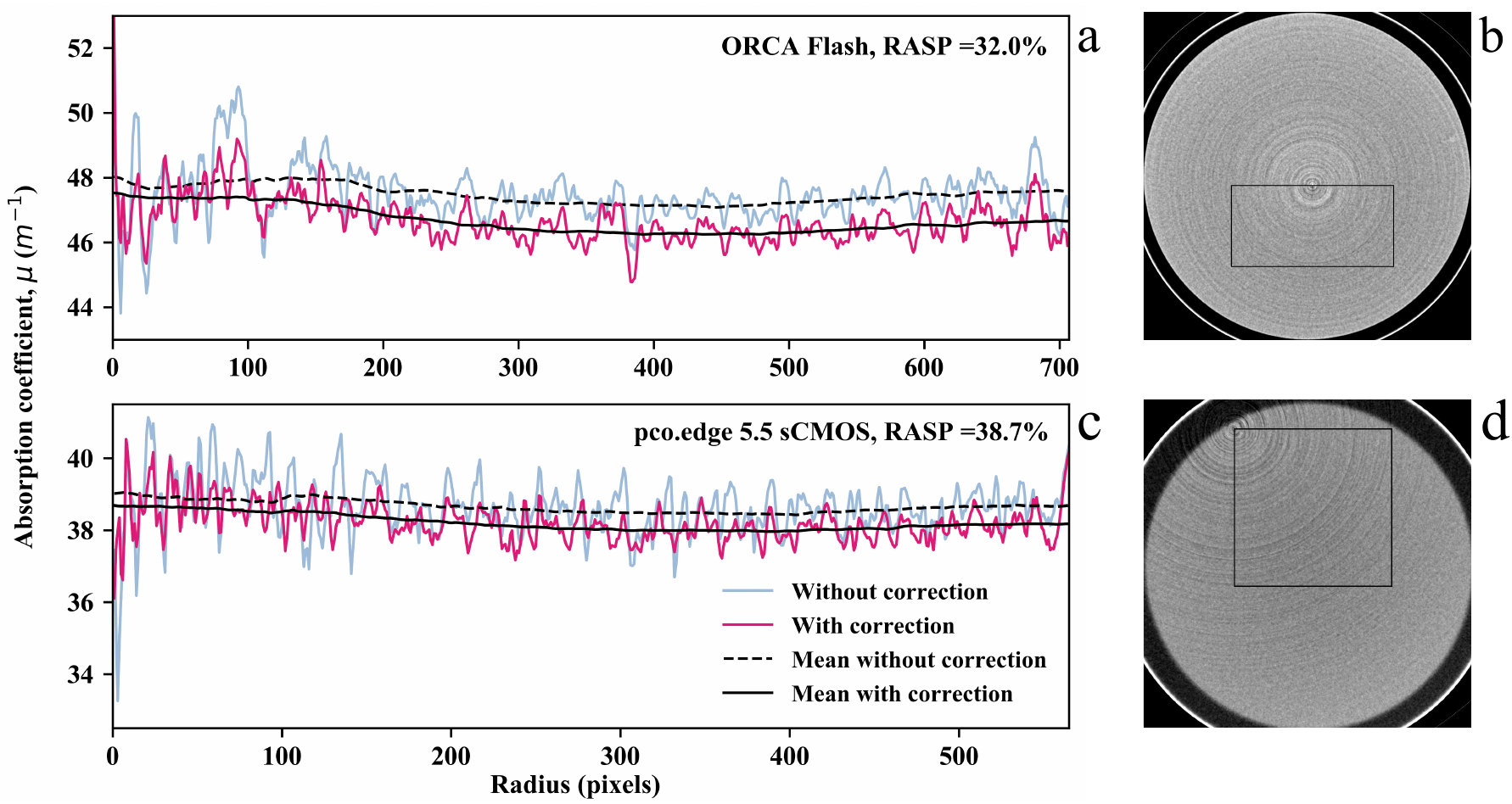}
    \caption{a) Azimuthally-averaged radial profiles for the sample-free region in Figs. \ref{fig:rings}(i) and \ref{fig:rings}(j). The decrease in the variance between the uncorrected and corrected images can be clearly seen, particularly at smaller radii, where the effects of the ring artifacts are strongest. b) The reconstruction region used for averaging (Hamamatsu ORCA Flash 4.0). Since the CT was acquired over 180$^{\circ}$, the artifacts for individual pixels occur only in the top or the bottom half of the image. c) Azimuthally-averaged radial profiles for the sample-free region in Figs. \ref{fig:rings}(k) and \ref{fig:rings}(l).  d) The reconstruction region used for averaging (pco.edge 5.5). The images in (b) and (d) are the uncorrected tomograms.}
  \label{fig:azavg}
\end{figure}

To better quantify the improvement, we define an image quality metric, the ring artifact suppression percentage (RASP), as the percentage reduction in the standard deviation ($\sigma$) of the azimuthally averaged radial profile from the COR in the sample-free region:
\begin{equation}
\makeatletter
\newcommand{\vast}{\bBigg@{6}}
\makeatother
\textrm{RASP} = \bigg(1 - \frac{\sigma_c}{\sigma_u}\bigg) \times 100\% = \vast(1 - \frac{\sqrt{\frac{1}{N}\sum\limits_{j=1}^{N} (x_{c,j} - \mu_{c,j})^2}}{\sqrt{\frac{1}{N}\sum\limits_{j=1}^{N} (x_{u,j} - \mu_{u,j})^2}}\vast) \times 100\%.
\end{equation}
Here, $c$ and $u$ denote the corrected and uncorrected images, respectively, $N$ is the total number of radial bins, $x_j$ is the azimuthally-averaged value within each bin, and $\mu_j$ is the mean value in each bin, determined by smoothing the radial profile with a median filter.
Figure \ref{fig:azavg}(a) shows the radial profiles and RASP obtained for the sample-free regions from Fig. \ref{fig:rings}(i)-(l), and Figs. \ref{fig:azavg}(b) and (c) show the regions used for averaging. The values obtained for these images are $\textrm{RASP} = 32.0\%$ for the ORCA Flash 4.0 used at SPring-8 and $\textrm{RASP} = 38.7\%$ for the pco.edge~5.5 used at the Australian Synchrotron. Averaging the RASP measurement over 100 consecutive slices within the sample-free regions give values of $\textrm{RASP} = 24.9 \pm 6.4\%$ and $\textrm{RASP} = 29.4 \pm 9.3\%$, respectively.

\begin{figure}[h!]
  \centering
      \includegraphics[width=1.0\textwidth]{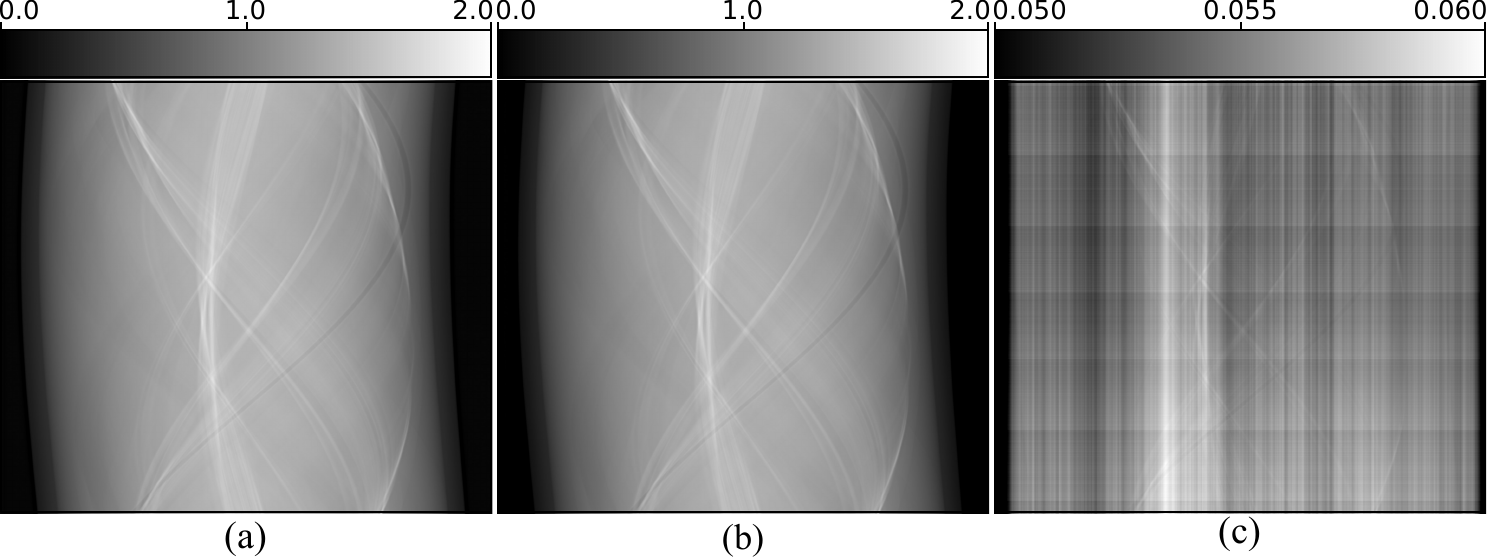}
    \caption{a) The uncorrected sinogram of the rabbit kitten head tomogram of Fig. \ref{fig:rings}(e). b) The sinogram of the same rabbit kitten head after the response correction has applied, corresponding to the tomogram of Fig. \ref{fig:rings}(f). Note that (a) and (b) look virtually identical, since ring artifacts that normally appear as stripes are overwhelmed by the signal from the skull. c) The difference between the sinograms in (a) and (b). The vertical stripes show the ring artifacts that have been removed. The horizontal stripes are caused by the additional correction that was applied to account for the time-variation in intensity due to beam injections (see step 8 in section 3).}
  \label{fig:diffsino}
\end{figure}

To further visualize these improvements, Fig. \ref{fig:diffsino} shows the change in ring artifacts seen between the sinograms corresponding to the reconstructed tomograms of Fig. \ref{fig:rings}(e) and (f), where the rings manifest as vertical stripes. The uncorrected and corrected sinograms of Fig. \ref{fig:diffsino}(a) and (b), respectively, demonstrate the extent to which these artifacts are overwhelmed by the signal from the skull, which makes sinogram-filtering techniques less effective. The difference image in Fig. \ref{fig:diffsino}(c) shows the rings (vertical stripes) that have been removed by the response correction described herein as well as the temporal variations in intensity (horizontal stripes) due to beam injections over the \textasciitilde 3-minute duration of the CT acquisition.

\section{Conclusions}

We have presented a simple correction to account for small spatial variations in detector response that is both easy to implement and highly effective at removing CT ring artifacts. This method is applicable for both absorption contrast and phase contrast imaging, and artifacts are suppressed to a level low enough to enhance even very low-contrast features, such as the soft-tissue boundaries within the brain. This correction requires just a single series of images to be acquired while sweeping the detector through the X-ray beam, and the data need only be acquired once for each set of experiments with a common detector configuration. In addition, the algorithm should be applicable to both laboratory and synchrotron experiments alike, with some modifications required for a polychromatic source. This correction is effective on its own but can also easily be combined with other existing ring artifact removal methods for further improvement. We have also defined a ring artifact suppression metric that can be used to assess the quality of any ring artifact removal technique, and we have used this metric to quantify the improvements achieved with our method.

\section*{Funding}

Research Training Program (RTP) Scholarship; ARC Future Fellowship (FT160100454); Veski Victorian Postdoctoral Research Fellowship (VPRF); German Excellence Initiative and European Union Seventh Framework Program (291763); NHMRC Development Grant (1093319); International Synchrotron Access Program (ISAP) (AS/IA153/10571).

\section*{Acknowledgements}

The authors would like to thank Michelle Croughan for her careful proofreading of this manuscript. The synchrotron radiation experiments were performed at the beamline BL20B2 of SPring-8 with the approval of the Japan Synchrotron Radiation Research Institute (JASRI) (Proposal No. 2017B0132). Additional research was undertaken on the Imaging and Medical Beamline (IMBL) (Proposal No. AS181/IMBL/12893) at the Australian Synchrotron, part of ANSTO.


\section*{Disclosures}

The authors declare that there are no conflicts of interest related to this article.

\end{document}